\documentclass[twocolumn, prl, superscriptaddress]{revtex4-2}
\usepackage{amssymb}
\usepackage{amsmath}
\usepackage{epsfig}
\usepackage{color}
\usepackage{graphics, graphicx}
\usepackage{bbold}
\usepackage{psfrag}
\usepackage{mathcomp}
\usepackage{subfigure}
\usepackage{verbatim}
\usepackage{color}
\usepackage[colorlinks,citecolor=blue]{hyperref}

\begin{document}
\title{Engineering Dissipative Quasicrystals}

\author{Tianyu Li}
\affiliation{CAS Key Laboratory of Quantum Information, University of Science and Technology of China, Hefei 230026, China}
\author{Yong-Sheng Zhang}
\email{yshzhang@ustc.edu.cn}
\affiliation{CAS Key Laboratory of Quantum Information, University of Science and Technology of China, Hefei 230026, China}
\affiliation{CAS Center For Excellence in Quantum Information and Quantum Physics, Hefei 230026, China}
\author{Wei Yi}
\email{wyiz@ustc.edu.cn}
\affiliation{CAS Key Laboratory of Quantum Information, University of Science and Technology of China, Hefei 230026, China}
\affiliation{CAS Center For Excellence in Quantum Information and Quantum Physics, Hefei 230026, China}

\begin{abstract}
We discuss the systematic engineering of quasicrystals in open quantum systems where
quasiperiodicity is introduced through purely dissipative processes. While the resulting short-time dynamics is governed by
non-Hermitian variants of the Aubry-Andr{\'e}-Harper model, we demonstrate how phases and phase transitions pertaining to the non-Hermitian quasicrystals fundamentally change the long-time, steady-state-approaching dynamics under the Lindblad master equation. Our schemes are based on an exact mapping between the eigenspectrum of the Liouvillian superoperator with that of the  non-Hermitian Hamiltonian, under the condition of quadratic fermionic systems subject to linear dissipation. Our work suggests a systematic route toward engineering exotic quantum dynamics in open systems, based on insights of non-Hermitian physics.
\end{abstract}

\maketitle

Famed for their aperiodic long-range order, quasicrystals exhibit distinct properties compared to materials with perfect lattice translational symmetry or under genuine disorder~\cite{harper, AA, sokoloff, roati, lahini, verbin}. In a periodic lattice potential, single-particle eigenstates are extended Bloch waves, which can become Anderson localized under disorder, constituting a fundamental mechanism for metal-insulator transitions in disordered systems~\cite{anderson, plee}. In a quasicrystal, such a disorder-driven localization also occurs but under significantly different conditions, as exemplified by the Aubry-Andr\'{e}-Harper (AAH) model in one dimension~\cite{AA}. Therein, the incommensurate lattice potentials drive a phase transition between the extended and the Anderson-localized states at critical quasiperiodicity, whereas such a transition is absent in one-dimensional disordered models as the Anderson localization takes over under infinitesimally small disorder. While the AAH model can also be mapped to a two-dimensional quantum Hall system~\cite{hofstadter, kraus, chen1}, thus acquiring an intriguing topological perspective, further generalizations such as adding off-diagonal quasiperiodic disorder or many-body interactions, would give rise to a number of exotic phases, epitomized by the  many-body localized~\cite{mbl1,mbl2,mbl3} or the critically-localized states~\cite{crit1,crit2,crit3}. As such, the AAH model and its variants have become appealing subjects for quantum simulations in systems such as photonics~\cite{verbin} and cold atoms~\cite{roati,crit3,gadway}, where both diagonal and off-diagonal quasiperiodic potentials can be engineered to investigate the rich physics of localization and topology.

Recently, amid the burgeoning interest in non-Hermitian physics, considerable attention has been drawn to the non-Hermitian generalization of the AAH model---an exemplary case of the so-called non-Hermitian quasicrystal~\cite{stefano1, yuce, liang, huan, harter, chen2, chen3, stefano2,liutong}.
On one hand, the interplay of non-Hermiticity and quasiperiodicity dramatically impacts properties such as the spectral symmetry and localization, giving rise to exotic phase transitions associated with spectral topology. On the other hand, the feasibility of simulating non-Hermitian Hamiltonians in the context of open systems renders the rich non-Hermitian phenomena experimentally relevant~\cite{nonHreview}.
Nevertheless, in quantum open systems, non-Hermitian Hamiltonians only govern the dynamics at short times, or, equivalently, under the condition of post-selection (conditional dynamics), while the long-time, unconditional dynamics is driven by the Liouvillian of the master equation~\cite{molmer, michael,nonHreview, weimer}. A fascinating question is whether the rich phase transitions of non-Hermitian quasicrystals would manifest themselves in the long-time dissipative dynamics?
In recent studies, it is shown that the non-Hermitian skin effect~\cite{ WZ1,WZ2,Budich,mcdonald,alvarez,murakami,ThomalePRB,fangchenskin,kawabataskin,Slager,yzsgbz,stefano,tianshu,lli}---an intriguing non-Hermitian phenomenon under which all eigenmodes become localized toward boundaries---can find its way into the long-time dynamics, giving rise to unique scaling behaviors and chiral damping patterns as the steady state is approached~\cite{wangzhong,yuzhenhua}. However, much less is known about the case with non-Hermitian quasicrystals~\cite{zhudamping}.

To address the question, here we explore the engineering of dissipative quasicrystals, for which key characters of non-Hermitian quasicrystals are systematically built into the dissipative quantum jump processes of a Lindblad master equation, such that dynamic transitions at long times are closely connected with the phase transitions of the non-Hermitian effective Hamiltonian.

We start by proving a general theorem, which establishes a direct mapping between the eigenspectrum of non-Hermitian effective Hamiltonians to the Liouvillian spectrum in quadratic open systems of fermions.
This provides a useful guide for engineering dissipative quasicrystals through quasiperiodic dissipation, which we demonstrate via three examples.
In all cases, the single-particle damping dynamics exhibits features that are directly related to a non-Hermitian quasicrystal. Beyond a critical quasiperiodic dissipation, the long-time single-particle correlation shows a localization transition, corresponding to the extended-localized phase transition in the non-Hermitian Hamiltonian. The quasiperiodic dissipation thus determines if the steady state is approached globally or in
localized patches.
Further, by adding non-reciprocal dissipation or quasiperiodic potentials, the localization transition acquires topological characters or coincides with a parity-time (PT)-symmetry-breaking point, giving rise to a wealth of dynamic signatures that originate from the non-Hermitian skin effect or the PT symmetry of the underlying non-Hermitian quasicrystal~\cite{chen3, stefano2}.

{\it From Liouvillian to Non-Hermiticity:---}
We consider a quadratic open quantum system of fermions in a lossy environment,
described by Lindbland master equation
\begin{align}
\frac{d\rho}{dt}=&-i[H_0,\rho]-\sum_{\mu}(L_\mu^\dag L_\mu\rho+\rho L^\dag_\mu L_\mu-2L_\mu\rho L_\mu^{\dagger}) \nonumber \\
=&\mathcal{L}\rho, \label{eq:lind}
\end{align}
where $\rho$ is the density matrix, the quadratic Hamiltonian $H_0=\sum_{i,j}^{n}h_{ij}c^\dagger_i c_j$, and the linear quantum jump operators $L_{\mu}=\sum_{i=1}^{n}l_{\mu,i}c_i$. Here $h_{ij}$ are the matrix elements of $H_0$, $c_i$ ($c_i^\dag$) is the fermionic annihilation (creation) operator of the $i$th mode, and $l_{\mu,i}$ is the corresponding mode-selective loss rate. From a quantum-trajectory perspective, the dynamics at short times is driven by the non-Hermitian Hamiltonian $H_{\text{eff}}=H_0-i\sum_{\mu}L_\mu^{\dagger}L_\mu$, in the absence of the quantum-jump terms $L_\mu\rho L_\mu^{\dagger}$~\cite{molmer, michael}. By contrast, the long-time dynamics is governed by the Liouvillian superoperator $\mathcal{L}$. The steady-state density matrix $\rho_s$ satisfies $\mathcal{L}\rho_s=0$.

Under the third quantization~\cite{prosen, okuma,lieu,dangel}, the Liouvillian superoperator can be formally written as $\mathcal{L}=\sum_{i} E_i \overline{b}^\dagger_i b_i$.
Here $E_i$ are the rapidities, and $b_i$, $\overline{b}_i$ are fermionic operators with $\{b_i, b_j\}=\{\overline b_i^\dagger, \overline b_j^\dagger\}=0$ and  $\{\overline b_i^\dagger, b_j\}=\delta_{ij}$.
While the steady state of the open system corresponds to the Fock vacuum of $b_i$, $E_i$ are essentially the eigenvalues of the normal-mode excitations above the steady state. The Liouvillian eigenspectrum $\lambda$ is generated by the rapidities through $\lambda=\sum_i E_i\nu_i$, with $\nu_i\in \{0,1\}$. Here $\lambda$
contains valuable information of the full density-matrix dynamics. Particularly, the Liouvillian gap, defined as the spectral gap between the steady state and the nearest normal mode, determines the scaling relation of the long-time dynamics: for a finite (vanishing) Liouvillian gap, the steady state is approached exponentially (algebraically)~\cite{wangzhong,zc}.

Alternatively, time evolution under the Lindblad equation (\ref{eq:lind}) is conveniently reflected in the dynamics of the single-particle correlation matrix $G(t)$, with matrix elements $G_{ij}(t)=\text{Tr}[c_i^\dagger c_j \rho(t)]$~\cite{wangzhong,diehl}. Here the damping dynamics of $G(t)$ is characterized by the damping matrix $X$, with $G(t)=e^{Xt}G(0)e^{X^\dag t}$ and $X_{ij}=ih_{ji}-\sum_\mu l^\ast_{\mu,j}l_{\mu,i}$. It is straightforward to show that $X=i h_{\text{eff}}^\ast$, where $(h_{\text{eff}})_{ij}$ are the corresponding matrix elements of $H_{\text{eff}}$. Thus, the eigenspectrum of the damping matrix $X$ and that of the non-Hermitian effective Hamiltonian contains the same information. But there is more to it.

\textbf{Theorem.} In a quadratic open system of fermions, if the eigenvalues of the damping matrix $X$ form the set $\{\lambda_n\}$, then the rapidities are $E_i=\{\lambda_n\}\cup \{\lambda_n^*\}$.

While the above theorem can be proved using the third quantization~\cite{supp}, the key message is that the eigenspectrum of $X$, and hence that of $H_{\text{eff}}$, contains all the information of the Liouvillian spectrum $\lambda$. This provides a close connection between the non-Hermitian effective Hamiltonian and the long-time dynamics. Particularly, the Liouvillian gap $\Delta$ can be easily constructed from eigenvalues of the damping matrix, with $\Delta=\min(|\text{Re} (\lambda_n)|)$~\cite{wangzhong}.

{\it Dissipative quasicrystals:---}
In the spirit of the theorem above, we introduce quasiperiodic dissipation to engineer non-Hermitian effective Hamiltonians, and show that long-time dynamic transitions in the full-fledged open-system dynamics are closely related to phase transitions of the underlying non-Hermitian quasicrystals.

We start by considering a tight-binding model with local quasiperiodic loss, with
\begin{align}
 H_0= \sum_n \left(t_1 c^\dagger_{n+1}c_n+ H.c.\right),
\end{align}
and the jump operators
\begin{align}
L^{(1)}_n=& \Gamma_nc_n, \label{eq:L1}
\end{align}
where $c_n$ ($c^\dag_n$) is the annihilation (creation) operator on site $n$, $t_1$ is the nearest-neighbor hopping rate, and the local loss rate $\Gamma_n= \sqrt{V[1+\sin(2\pi \alpha n)]}$, with $\alpha=(\sqrt{5}-1)/2$ and $V$ being the strength of the quasiperiodic dissipation. The corresponding non-Hermitian Hamiltonian is a generalized non-Hermitian AAH model with quasiperiodic on-site loss, which, apart from a global loss, is written as
\begin{align}
H_1=\sum_n \left(t_1 c^\dagger_{n+1}c_n+ H.c.\right)-\sum_n iV \sin(2\pi\alpha n) c^\dagger_nc_n. \label{eq:H1}
\end{align}
Invoking the Thouless formula~\cite{thouless,songz,supp}, the inverse localization length can be analytically derived as $\eta=\log(V/2t_1)$.
A localization-delocalization transition thus occurs at $V/2t_1=1$, where $\eta$ vanishes.

\begin{figure}[tbp]
\centering
\includegraphics[width=8cm]{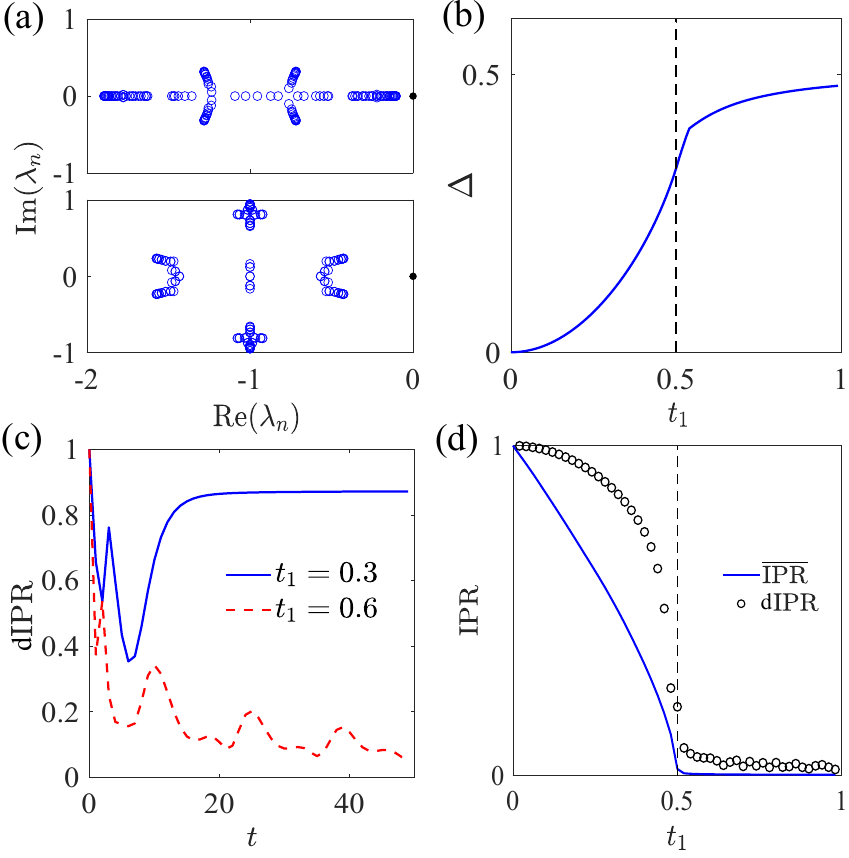}
\caption{
Localization transition of $H_1$ and its manifestation in open-system dynamics.
(a) Eigenspectra of the damping matrix $X$ on the complex plane, with $t_1=0.3$ (top) and $t_1=0.6$ (bottom), respective. The eigenvalue of steady state ($\lambda_s=0$) is indicated by the black dot. We take a lattice with $N=144$ sites.
(b) The Liouvillian gap $\Delta$ as a function of $t_1$.
(c) Dynamic evolution of $\text{dIPR}(t)$.
(d) Comparison between $\overline{\text{IPR}}$ from the non-Hermitian Hamiltonian $H_1$ (blue line), and $\text{dIPR}(t=100)$ from full open-system dynamics (black symbol). We fix $V=1$ for all calculations, and  initialize the system in a fully localized single-particle state in the bulk.
}\label{fig:fig2}
\end{figure}

\begin{figure}[tbp]
\centering
\includegraphics[width=8cm]{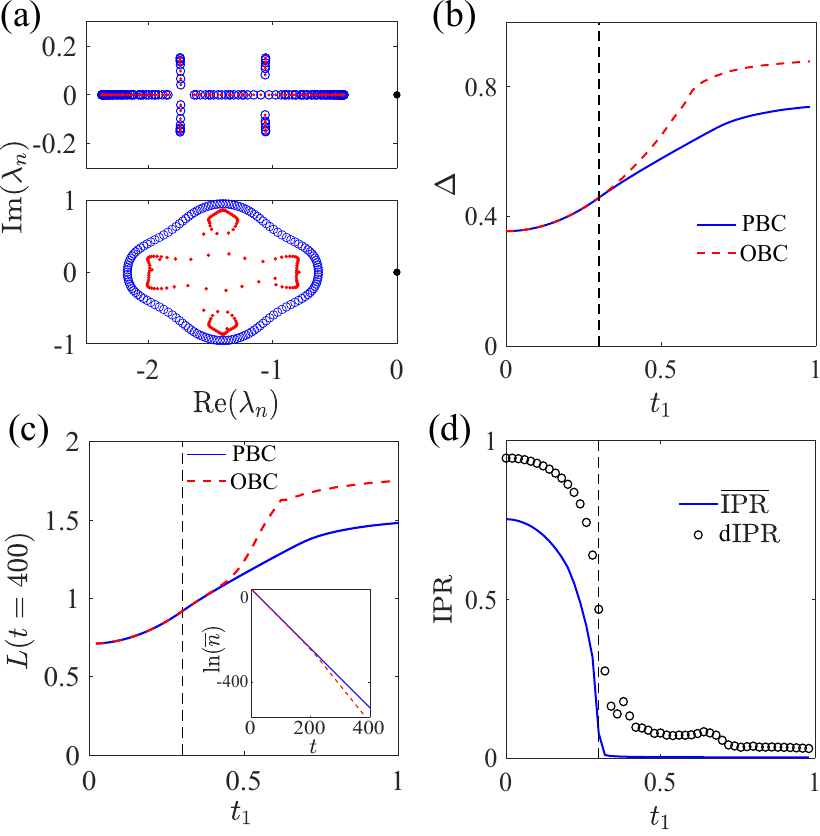}
\caption{
Dissipative quasicrystal corresponding to $H_2$.
(a) Eigenspectra of the damping matrix $X$, with $t_1=0.25$ (top) and $t_1=0.6$ (bottom).
Blue (red) symbols show calculations under PBC (OBC). We take a total lattice site of $N=144$.
(b) The Liouvillian gap $\Delta$ as a function of $t_1$.
(c) Damping rate $L(t=400)$ as a function of $t_1$. (Inset): the numerically simulated mean particle number $\overline n(t)$ with $t_1=0.6$, under different boundary conditions.
(d) Comparison between $\overline{\text{IPR}}$ for $H_2$ (blue line), and $\text{dIPR}(t=100)$ from full open-system dynamics (black symbol). The same initial condition is taken
as that in Fig.~\ref{fig:fig2}, and we fix $V=1$, $\varphi=0$, and $\gamma=0.2$ for all calculations. The phase transition is located at $t_1=0.3$.
 }\label{fig:fig3}
\end{figure}

The localization transition does not impact the Liouvillian gap $\Delta$ in a qualitative way, albeit $\Delta$ is generally smaller in the localized state (Fig.~\ref{fig:fig2}(a)(b)).
More relevantly, the phase transition can be characterized by the mean inverse participation ratio, defined as  $\overline{\text{IPR}}=\frac{1}{N}\displaystyle{\sum_{i,n}}|\psi_{i,n}|^4$~\cite{crit2}, where $\psi_{i,n}$ is the support of the $i$th eigen wavefunction of $H_1$ on lattice site $n$. As shown in Fig.~\ref{fig:fig2}(d), the calculated $\overline{\text{IPR}}$ is vanishingly small for $t_1>V/2$, indicating an extended state; while it becomes finite for $t_1<V/2$, indicating localization.

To see the impact of the localization transition on the long-time dynamics, we focus on the dynamic inverse participation ratio, $\text{dIPR}(t)=\sum_n G_{nn}(t)^2/[\sum_n G_{nn}(t)]^2$~\cite{crit3}. Here $G_{nn}$ is essentially the on-site population. As shown in Fig.~\ref{fig:fig2}(c)(d), while $\text{dIPR}$ features oscillations at short times, it shows distinct behaviors at long times for the two phases. A dynamic phase transition point can be identified from the long-time $\text{dIPR}(t)$, which agrees with the localization transition of the non-Hermitian Hamiltonian $H_1$. Physically, the dynamic transition indicates different ways the steady state (the vacuum state $|0\rangle$ here) is approached---globally (for $t_1>V/2$) or in localized patches (for $t_1<V/2$).

We now move on to the second case, by further introducing non-reciprocal dissipation processes, with additional jump operators~\cite{wangzhong}
\begin{align}
L^{(2)}_{n}=&\sqrt{\gamma}(c_n-ic_{n+1}),
\end{align}
where $\gamma$ is the loss rate. Apart from a global loss term, the corresponding non-Hermitian effective Hamiltonian is
\begin{align}
H_{2}=&\sum_n \Big[J e^\beta c^\dagger_{n+1}c_n+ Je^{-\beta} c^\dagger_{n}c_{n+1}\nonumber\\
 &- iV \sin(2\pi\alpha n+\varphi) c^\dagger_nc_n\Big],
\end{align}
where $Je^{\pm \beta}=t_1\pm\gamma$, where $\varphi$ is a phase factor. Jump operators $L^{(2)}_n$ give rise to a non-reciprocal nearest-neighbor hopping in the non-Hermitian Hamiltonian, which in turn induces the non-Hermitian skin effect under open boundary conditions (OBCs). Remarkably, the onset of the non-Hermitian skin effect coincides with the localization transition, while the transition point is shifted due to the competition between the non-Hermitian skin effect and the Anderson localization~\cite{chen3}.

\begin{figure}[tbp]
\centering
\includegraphics[width=8cm]{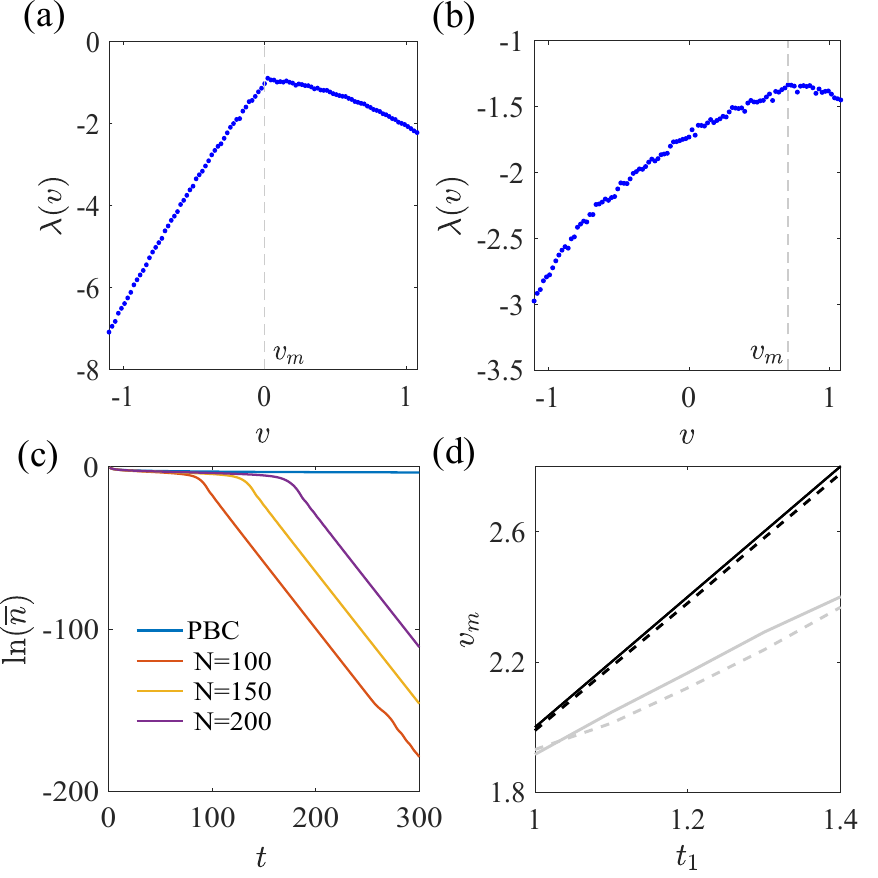}
\caption{
Lyapunov exponent and chiral damping.
(a)(b) Lyapunov exponent $\lambda(v)$ in a full open-system dynamics, with $t_1=0.25$ and $V=1$ (a), and
$t_1=0.6$ and $V=1$ (b).
The initialization is the same as that in Fig.~\ref{fig:fig2}.
(c) Damping of the mean particle number under different boundary conditions, with $V=0$ and different lattice lengths. The overall damping behavior changes from algebraic (under PBC), to exponential (under OBC) when the chiral damping wavefront traverses the lattice. The propagation speed of the wavefront is estimated by dividing $N$ with the critical time of the transition.
(d) Comparing the drift velocity $v_m$ (solid) where $\lambda(v)$ peaks, with the speed (dashed) of the chiral-damping wavefront. The black lines (solid and dashed) indicate $V=0$, and the gray ones $V=0.4$.
For (c)(d), we initialize the system in a fully filled state. We fix $\gamma = 0.2$ for all calculations.
For the gray solid line in (d), we take an average over $20$ randomly chosen $\varphi$ in the range of $[0,2\pi)$, while $\varphi=0$ otherwise.
 }\label{fig:fig4}
\end{figure}

\begin{figure}[tbp]
\centering
\includegraphics[width=8cm]{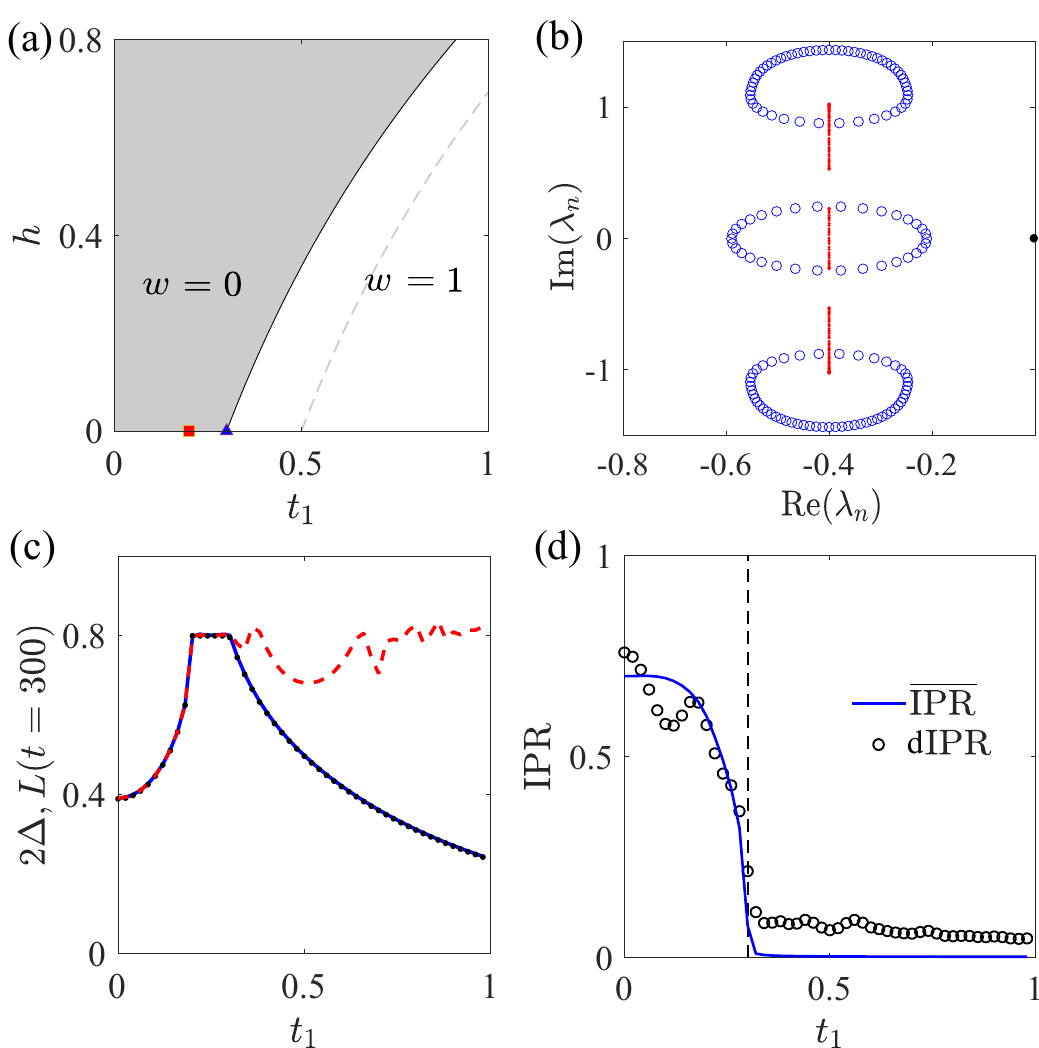}
\caption{
(a) Phase diagram of the spectral winding number $\omega$, for $H_3$ with $\gamma=0.2$ (solid black) and $\gamma=0$ (dashed gray), respectively.
The red square and blue triangle indicates the exceptional point of $H_3$, with $t_1=0.2$ and $t_1=0.3$, respectively.
(b) The PBC eigenspectra of $X$ associated with $H_3$, with $t_1=0.25$ (red) and $t_1=0.6$ (blue), respectively.
(c) Damping rate $L(t)$ as a function of $t_1$ under PBC (blue) and OBC (red), respectively. We also plot $2\Delta$ (black dots) for comparison.
(d) Comparison between $\overline{\text{IPR}}$ from $H_3$ (blue line), and $\text{dIPR}(t=30)$ from full open-system dynamics (black symbol). The same initial condition is taken
as that in Fig.~\ref{fig:fig2}.
We take $\gamma=0.2$, $h=0$, and a lattice size of $N=144$ in (b)(c), and fix $V=1$ for all calculations.
}
\label{fig:fig5}
\end{figure}

To derive the transition point, we notice that $H_1$ and $H_2$ are related by a similarity transformation under OBC, with $H_2=SH_1S^{-1}$ and $S=\text{diag}(e^\beta, e^{2\beta},..., e^{N\beta})$. The localized eigen wavefunctions of $H_2$ thus behave as $e^{-(\eta\pm\beta) |n-n_0|}$, where $n_0$ is the localization center. As such, the localization transition occurs at $V/2J=e^{\beta}$~\cite{chen3}.

In Fig.~\ref{fig:fig3}, we plot the eigenspectrum of the damping matrix for the second case. When $H_2$ is in the localized phase, the eigenspectrum is the same under OBC and under a periodic boundary condition (PBC), indicating the absence of the non-Hermitian skin effect (Fig.~\ref{fig:fig3}(a)(b)). By contrast, in the extended phase, a spectral loop emerges under PBC, whereas it shrinks to open spectral arcs under OBC. While the transition point is numerically confirmed by calculating $\overline{\text{IPR}}$ (Fig.~\ref{fig:fig3}(d)), the transition itself now acquires a spectral topological nature. This is seen by introducing a spectral winding number	~\cite{ueda,chen3}
\begin{align}
\omega=\lim_{N\rightarrow \infty}\frac{1}{2\pi i }\int^{2\pi}_{0}d\Phi\frac{\partial_\Phi \det H(\Phi)}{\det H(\Phi)},
\end{align}
where $\Phi$ is an auxiliary magnetic flux threaded through the one-dimensional ring under PBC, with $H(\Phi)=H_2 + Je^{\beta}e^{-i\Phi} c^\dagger_{1}c_N+Je^{-\beta}e^{i\Phi} c^\dagger_{N}c_{1}$.
Consistent with the PBC spectra in Fig.~\ref{fig:fig3}(a), the winding number vanishes in the localized phase, but becomes unity in the extended phase, suggesting the presence of non-Hermitian skin effect~\cite{fangchenskin,kawabataskin}.

Similar to the previous case, the localization characters of $H_2$ carry over to the open-system dynamics, reflected in the $\text{dIPR}$ at long times (Fig.~\ref{fig:fig3}(d)). Additionally, the presence of the non-Hermitian skin effect in the Liouvillian eigenspectrum suggests the long-time damping dynamics should have distinct behaviors under different boundary conditions. This is directly shown in Fig.~\ref{fig:fig3}(c), where the damping rate $L(t)=-\frac{d}{dt}\ln \overline{n}(t)$ at long times agrees well with $L(t=\infty)=2\Delta$, both sensitive to the boundary condition in the extended phase where the non-Hermitian skin effect is present.
Here the mean occupation number $\overline{n}(t)$ is defined as $\overline{n}(t)=\frac{1}{N}\sum_x G_{xx}(t)$.

A prominent signature of the non-Hermitian skin effect lies in the Lyapunov exponent of the open-system dynamics, defined as $\lambda(v)=\displaystyle{\lim_{t\to \infty}}\frac{1}{t}\log|G_{x=vt,x=vt}(t)|$, where $v$ is the drift velocity.
As shown in Fig.~\ref{fig:fig4}(a)(b), $\lambda(v)$ peaks at $v_m\neq 0$ ($v_m=0$) in the presence (absence) of the non-Hermitian skin effect.
The behavior is the same as that of the Lyapunov exponent under a non-Hermitian Hamiltonian~\cite{stefano,xueexp},
defined
as $\lambda'(v)=\displaystyle{\lim_{t\to \infty}}\frac{1}{t}\log |\psi_{x=vt}(t)|$ ($\psi_x(t)$ is the time-evolved wavefunction component on site $x$).
In fact, it can be explicitly proved that $\lambda(v)=-2(V+2\gamma)+2\lambda'(v)$~\cite{supp}.

Here in open systems, $v_m$ further provides a quantitative measure for the chiral damping under OBC, where a sharp wavefront traverses the lattice in a chiral fashion, a manifestation of the non-Hermitian skin effect of the Liouvillian~\cite{wangzhong}. As shown in Fig.~\ref{fig:fig4}(c)(d), $v_m$ exactly matches the speed of the wavefront propagation in the absence of quasiperiodicity ($V=0$). While it decreases under the quasiperiodic dissipation (finite $V$), it is still consistent with the latter, averaged over the phase $\varphi$ of the quasiperiodic disorder.
Intuitively, a finite peak drift velocity in $\lambda(v)$ and the wavefront propagation in chiral damping both reflect the persistent bulk current underlying the non-Hermitian skin effect in open-system quantum dynamics.

For the last example, we introduce a quasiperiodic on-site potential $V'\cosh h \cos(2\pi\alpha n)$, where $h$ is a real parameter. We now parameterize the quasiperiodic loss rate in Eq.~(\ref{eq:L1}) as $V=V'\sinh h$.
The non-Hermitian effective Hamiltonian becomes
\begin{align}
H_3=&\sum_n \Big[Je^{\beta} c^\dagger_{n+1}c_n+Je^{-\beta} c^\dagger_{n}c_{n+1} \nonumber\\
&+ V'\cos(2\pi\alpha n+ih) c^\dagger_nc_n\Big],\label{eq:H3}
\end{align}
whose localization phase boundary can be analytically derived as $V'/2J=e^{|\beta|-|h|}$~\cite{chenshuprb}.

For $h=0$, the localization transition at $t_1=0.3$ coincides with a PT-symmetry breaking transition (Fig.~\ref{fig:fig5}(a)(b)).
Therein, three different phase transitions occur simultaneously at $t_1=0.3$ and $h=0$~\cite{chen3}.
While localization and non-Hermitian skin effects leave similar dynamic signatures as the previous cases, the PT-symmetry breaking transition modifies the Liouvillian gap (Fig.~\ref{fig:fig5}(b)(c)), and affects the long-time damping rate $L(t)$ (Fig.~\ref{fig:fig5}(c)).
For $h\neq 0$, $H_3$ is no longer PT symmetric, but the localization transition still acquires a spectral topological nature, and simultaneously mark the appearance (in the extended state) or disppearance (in the localized state) of the non-Hermitian skin effect.

{\it Discussion:---}
By engineering quantum jump operators in an open dissipative system, we demonstrate how exotic phases and phase transitions in non-Hermitian quasicrystals
induce transitions in the full open-system quantum dynamics, characterized by the change of spatial or temporal patterns the steady state is approached.
Our systematic construction of the dissipative quasicrystal is aided by the close connection between non-Hermitian phenomena and dynamics in quadratic Fermi open systems. Such an understanding would offer the blueprint, in the form of non-Hermitian Hamiltonians, for conceiving interesting quantum open systems.


\begin{acknowledgments}
{\it Acknowledgments.---}This work has been supported by the Natural Science Foundation of China (Grant Nos. 11974331, 92065113) and the National Key R\&D Program (Grant Nos. 2016YFA0301700, 2017YFA0304100).
\end{acknowledgments}
\bibliographystyle{apsrev4-1}

\clearpage
\begin{widetext}
\appendix

\renewcommand{\thesection}{\Alph{section}}
\renewcommand{\thefigure}{S\arabic{figure}}
\renewcommand{\thetable}{S\Roman{table}}
\setcounter{figure}{0}
\renewcommand{\theequation}{S\arabic{equation}}
\setcounter{equation}{0}

\section{Supplemental Materials}

Here we provide details on the proof of the theorem in the main text, as well as discussions on
the long-time damping dynamics, the characterization of the localization length and the Lyapunov exponent.

\section{Damping dynamics}

We consider a quadratic open Fermi system, where the density matrix $\rho$ evolves under the Linblad maseter equation
\begin{align}
\frac{d\rho}{dt}=&-i[H,\rho]+\sum_{\mu}(2L_\mu\rho L_\mu^{\dagger}-\{L_\mu^\dagger L_\mu,\rho\}) \\\nonumber
=&\mathcal{L}\rho,
\end{align}
with
\begin{align}
 H=\sum_{i,j}^{n}c^\dagger_ih_{ij}c_j, \quad  L_{\mu}=\sum_{i=1}^{n}l_{\mu,i}c_i . \label{eq:H}
\end{align}

In the single-particle subspace, the Lindblad equation has the solution
\begin{align}
\rho(t)=e^{-iH_{\text{eff}}t}\rho_0 e^{iH^\dagger_{\text{eff}}t}+\{1-\text{Tr}[e^{-iH_\text{eff}t}\rho_0 e^{iH^\dagger_\text{eff}t}]\}|0\rangle\langle 0|,
\end{align}
where $\rho_0$ is the initial density matrix, and $H_{\text{eff}}=H_0-i\sum_{\mu}L_\mu^{\dagger}L_\mu$.
Note that $L_\mu \rho_0 L^\dagger_\mu=\text{Tr}(L_\mu \rho_0 L^\dagger_\mu)|0\rangle\langle 0|$, and $L_\mu|0\rangle=0$, with the steady-state density matrix given by $|0\rangle\langle 0|$.

It follows that, starting from a single-particle subspace, the particle number dynamics is given by $n(t)=\text{Tr}(e^{-iH_{\text{eff}}t}\rho_0 e^{iH^\dagger_{\text{eff}}t})$, scaling exponentially for $t\rightarrow \infty$~\cite{wangzhong}. Here we have used the relation $X=ih^*_{\text{eff}}$. Such a long-time damping behavior is consistent with results in the main text.

\section{Connecting the eignsspectrum of the damping matrix and the Liouvillian}

We rewrite Eq~(\ref{eq:H}) in the Majorana presentation~\cite{prosen}
 \begin{align}
 H=\sum_{i,j}^{2n}\gamma_iH^M_{ij}\gamma_j, \quad  L_\mu=\sum_{i=1}^{2n}l^M_{\mu,i}\gamma_i,
\end{align}
where $\gamma_i$ is the operator of Majorana fermions, with $\gamma_{2m-1}=c_m+c_m^\dagger, \gamma_{2m}=i(c_m-c_m^\dagger)$, and $\{\gamma_i, \gamma_j\}=2\delta_{ij}$. $H^M$ is chosen to be anti-Hermitian, $(H^M)^T=-H^M$. Rewriting $h$ as $h=h_{r}+ih_{i}$, where $h_r$ and $h_i$ are real matrices, we have $H^M=\frac{1}{4}(h_{r}\otimes \sigma_y+ih_{i}\otimes \mathbf{1})$. Here $\mathbf{1}$ is the $2\times 2$ identity matrix, and $\sigma_y$ is the Pauli matrix.
Under the definition $M_{ij}=\sum_\mu l_{\mu,i}^* l_{\mu,j}$, $M^M_{ij}=\sum_\mu (l^M_{\mu,i})^* l^M_{\mu,j}$, we have $M^M=\frac{1}{4} M\otimes (\mathbf{1}+\sigma_y)$.

Invoking third quantization formalism, the Liouvillian superoperator can be represented by $2N$ fermionic operators~\cite{okuma}.
\begin{align}
\mathcal{L}=\frac{2}{i}(\mathbf{f}^\dagger \quad \mathbf{f})
\begin{pmatrix}
-Z^{T}& Y \\
0 & Z
\end{pmatrix}
\begin{pmatrix}
\mathbf{f}\\
\mathbf{f}^\dagger
\end{pmatrix},
\end{align}
where $Z=H^M+i \text{Re} (M^M)^T, Y=2 \text{Im} (M^M)^T$, and $\mathbf{f}=(f_1,...,f_{2N})$.

Further, we have
\begin{align}
\mathcal{L}=\sum_{i=1}^{2n}E_i \overline{b}^\dagger_i b_i,
\end{align}
where $b_i$ and $\overline{b}_i$ are fermionic operators with $\{b_i, b_j\}=\{\overline b_i^\dagger, \overline b_j^\dagger\}=0$, and $\{\overline b_i^\dagger, b_j\}=\delta_{ij}$. $\{E_i\}$ are the eigenvalues of $4iZ$, known as the rapidities. The rapidities can generate all the $2^{2N}$ Liouvillian eigenvalues through
\begin{align}
\lambda=\sum_{i=1}^{2n}E_i\nu_i
\end{align}
where $\nu_i\in \{0,1\}$.

\textbf{Lemma.} If $A$, $B$ are two $n\times n$ real matrices, and the eigenvalues of $A+iB$ form the set $\{\lambda_n\}$, then the eigenvalues of  $\mathbf{1}\otimes A+i\sigma_y\otimes B$ form the set $\{\lambda_n\}\cup \{\lambda_n^*\}$.

Proof: Since
$\mathbf{1}\otimes A+i\sigma_y\otimes B$= $\begin{pmatrix}
A& B \\
-B &A
\end{pmatrix}$,
we have
det$\begin{pmatrix}
A-\lambda& B \\
-B &A-\lambda
\end{pmatrix}$=$\det(A+iB-\lambda)\det(A-iB-\lambda)=0$.
Since $A, B$ are real matrices, we have $A-iB=(A+iB)^*$. Therefore, if $\{\lambda_n\}$ are the eigenvalues of $A+iB$, $\{\lambda_n^*\}$ are the eigenvales of $A-iB$.

\textbf{Theorem.} In a quadratic open system of fermions, if the eigenvalues of the damping matrix $X$ form the set $\{\lambda_n\}$, then the rapidities are $E_i=\{\lambda_n\}\cup \{\lambda_n^*\}$.

Proof:
Rewriting $M$ as $M=M_{r}+iM_{i}$, where $M_{r}, M_{i}$ are real matrices, we have $M^M=\frac{1}{4}(M_{r}+iM_{i})\otimes (\mathbf{1}+\sigma_y)$.

It follows that
\begin{align}
Z&=H^M+i\text{Re} (M^M)^T=\frac{1}{4}(h_{r}\otimes \sigma_y+ih_{i}\otimes \mathbf{1})+i\frac{1}{4}(M_{r}^T\otimes \mathbf{1}-iM_{i}^T\otimes \sigma_y)\nonumber\\
&=\frac{1}{4}(h_{r}+M_{i}^T)\otimes \sigma_y+i\frac{1}{4}(h_{i}+M_{r}^T)\otimes \mathbf{1}.
\end{align}
Therefore, $4iZ=(h_{r}+M_{i}^T)\otimes i\sigma_y-(h_{i}+M_{r}^T)\otimes \mathbf{1}$. According to the Lemma,  the eigenvalue of $4iZ$ are the union of the eigenvalues of $X$ and $X^*$.


\section{The localization length}

Consider a Hamiltonian $H$ in the lattice-site basis $\{|n\rangle\}$
\begin{align}
H=
\begin{pmatrix}
V_1& J  \\
J &V_2&J \\
 & \ddots& \ddots&\ddots\\
 &&J &V_N\\
\end{pmatrix},\label{eq:H'}
\end{align}
 where $V_n=-iV\sin(2\pi\alpha n)$.

While the Thouless formula is originally derived for a Hermitian Hamiltonian~\cite{thouless}, it can be generalized to a broad class of non-Hermitian Hamiltonians and is applicable here.
Specifically, since $H$ is symmetric, it can be decomposed as $H=\sum_n\lambda_n |\psi_n\rangle\langle \psi_n^*|$,
where $H|\psi_n\rangle=\lambda_n|\psi_n\rangle$, $\langle\psi_n^*|H=\lambda_n\langle\psi_n^*|$, and $\langle \psi_m^*| \psi_n\rangle=\delta_{mn}$.
Suppose $G(\lambda)$ is the inverse matrix of $H-\lambda$, which has the decomposition $G(\lambda)=\sum_n1/(\lambda_n-\lambda) |\psi_n\rangle\langle \psi_n^*|$.
The residue of $G_{1N}(\lambda)$ at $\lambda_n$ is then given by $a_1^na_N^n$, where $a_m^n$ is the $m$th component of $ |\psi_n\rangle$. Taking advantage of the
tridiagonal property of $G(\lambda)$, the matrix element $G_{1N}(\lambda)$ can be derived as
\begin{align}
G_{1N}(\lambda)=&J^{N-1}/\det(H-\lambda)\nonumber\\
=& J^{N-1}/\prod_n(\lambda_n-\lambda),
\end{align}
where the residue of $G_{1N}(\lambda)$ at $\lambda_n$ is found to be $J^{N-1}/\prod_{m\neq n}(\lambda_m-\lambda_n)$. Thus, we have $a_1^na_N^n=J^{N-1}/\prod_{m\neq n}(\lambda_m-\lambda_n)$. The inverse localization length is then
\begin{align}
\eta(\lambda_n)=\lim_{N\rightarrow \infty}\frac{-\ln|a_N^n a_1^n|}{N-1}=\lim_{N\rightarrow \infty}1/(N-1)\sum_{m\neq n}\ln|\lambda_m-\lambda_n|-\log(J),
\label{eq:thouless}
\end{align}
which is exactly the Thouless formula.

Taking the Fourier transform~\cite{sokoloff,songz}
\begin{align}
|n\rangle=\sum_k(-i)^ki^ne^{i2\pi\alpha nk}|k\rangle,
\end{align}
we get the dual model
\begin{align}
iH=\sum_k(\frac{V}{2}|k\rangle\langle k+1|+h.c.)-\sum_k i2J\sin(2\pi\alpha k)|k\rangle\langle k|.
\end{align}

Note that the dual model must have the same eigenvalues with Eq.~(\ref{eq:H'}) except for a factor $i$. So the inverse localization length of the dual model is $\eta'(\lambda_n)=\lim_{N\rightarrow \infty}1/(N-1)\sum_{m\neq n}\ln|\lambda_m-\lambda_n|-\log(V/2)$. We then have $\eta(\lambda_n)=\eta'(\lambda_n)+\log(V/2J)$. When the eigenstates are exponentially localized in the lattice-site basis, they must be delocalized in the dual model. We therefore have $\eta'(\lambda_n)=0$ and  $\eta(\lambda_n)=\log(V/2J)$ for $V>2J$ ($\eta>0$ for both models). Similarly, we have $\eta'(\lambda_n)=\log(2J/V)$ for $V<2J$. Note that the inverse localization length is independent of the eigenvalues $\lambda_n$. The phase transition point is therefore $V=2J$~\cite{sokoloff}.


%

\section{The Lyapunov exponent}
In the second case of the main text, we have $H_{\text{eff}}=H_2-i(V+2\gamma)$, hence
\begin{align}
G(t)=e^{Xt}G(0) e^{X^\dagger t}=e^{-2(V+2\gamma)t}(e^{-iH_2t}G^*(0) e^{iH_2^\dagger t})^*,
\end{align}
where we have used $X=ih^*_{\text{eff}}$. We take the initial state of the system $c^\dagger_{N/2}|0\rangle$, so formally one has $G(0)=|\frac{N}{2}\rangle\langle \frac{N}{2}|$. It follows that $G(t)=e^{-2(V+2\gamma)t}(e^{-iH_2t}|\frac{N}{2}\rangle\langle \frac{N}{2}| e^{H_2^\dagger t})^*$. In the non-Hermitian Hamiltonian, the Lyapunov exponent is defined as $\lambda'(v)=\displaystyle{\lim_{t\to \infty}}\frac{1}{t}\log|\psi_{x=vt}(t)|$, where $v$ is the drift velocity~\cite{stefano} and $\psi_x$ is the
support of the time-evolved wavefunction on site $x$.
The Lyapunov exponent of the open-system $\lambda(v)$ is related to that of the non-Hermitian Hamiltonian, through $\lambda(v)=\displaystyle{\lim_{t\to \infty}}\frac{1}{t}\log|G_{x=vt,x=vt}(t)|= -2(V+2\gamma)+2\lambda'(v)$. Thus the peak in $\lambda(v)$ is consistent with that of $\lambda'(v)$, both reflect the speed of wavefront in the chiral damping of an open system.

On the other hand, for the case with $V=0$, the dispersion in the quasimomentum $k$ space is $E(k)=2t_1 \cos k-i2\gamma \sin k$.
The eigenvalue with the largest imaginary component is taken at $k_0=-\pi/2$. Following Ref.~\cite{stefano}, the dominant drift velocity, where the Lyapunov exponent peaks, is given by $v_m=\frac{d}{dk}\text{Re}E(k)\Big|_{k=k_0}=2t_1$.  This is consistent with the numerical results in Fig.~3(d) of the main text. By contrast, in the presence of quasiperiodicity (under finite $V$), $k$ is no longer a good quantum number. The dominant $v_m$ becomes smaller, but is consistent with the speed of wavefronts in chiral damping, averaged over different phases of the quasiperiodic disorder (see gray curves in Fig.~3(d)).

\end{widetext}

\end{document}